\documentclass{IEEEtran}

\usepackage{xcolor,soul,framed} %,caption

\colorlet{shadecolor}{yellow}
\usepackage[pdftex]{graphicx}
\graphicspath{{../pdf/}{../jpeg/}}
\DeclareGraphicsExtensions{.pdf,.jpeg,.png}

\usepackage[cmex10]{amsmath}
\usepackage{array}
\usepackage[numbers,sort&compress]{natbib}
\usepackage{mdwmath}
\usepackage{mdwtab}
\usepackage{eqparbox}
\usepackage{url}
\usepackage{hyperref}
\usepackage{subfigure}
\usepackage{soul}
\soulregister\cite7 
\soulregister\ref7
\usepackage{lineno}

\def\BibTeX{{\rm B\kern-.05em{\sc i\kern-.025em b}\kern-.08em
T\kern-.1667em\lower.7ex\hbox{E}\kern-.125emX}}
\begin{document}
%\linenumbers

\title{Development of silicon interposer: towards an ultralow radioactivity background photodetector system}

\author{\IEEEauthorblockN{Haibo Yang\IEEEauthorrefmark{1},
Qidong Wang\IEEEauthorrefmark{1},
Guofu Cao\IEEEauthorrefmark{2}\IEEEauthorrefmark{3},
Kali M. Melby\IEEEauthorrefmark{4},
Khadouja Harouaka\IEEEauthorrefmark{4},
Isaac J. Arnquist\IEEEauthorrefmark{4},
Fengwei Dai\IEEEauthorrefmark{1},
Liqiang Cao\IEEEauthorrefmark{1},
Liangjian Wen\IEEEauthorrefmark{2}}

\IEEEauthorblockA{\IEEEauthorrefmark{1}Institute of Microelectronics, Chinese Academy of Sciences, Beijing 100029, China}

\IEEEauthorblockA{\IEEEauthorrefmark{2}Institute of High Energy Physics,Chinese Academy of Sciences, Beijing 100049, China}

\IEEEauthorblockA{\IEEEauthorrefmark{3}University of Chinese Academy of Sciences, Beijing 100049, China}

\IEEEauthorblockA{\IEEEauthorrefmark{4}Pacific Northwest National Laboratory, Richland, WA 99352, USA}

\thanks{Corresponding author: Qidong Wang (wangqidong@ime.ac.cn).}
\thanks{Corresponding author: Guofu Cao (caogf@ihep.ac.cn).}
\thanks{}}

\maketitle

\begin{abstract}
It is of great importance to develop a photodetector system with an ultralow radioactivity background in rare event searches. Silicon photomultipliers (SiPMs) and application-specific integrated circuits (ASICs) are two ideal candidates for low background photosensors and readout electronics, respectively, because they are mainly composed of silicon, which can achieve good radio-purity without considerable extra effort. However, interposers, used to provide mechanical support and signal routes between the photosensor and the electronics, are a bottleneck in building ultralow background photodetectors. Silicon and quartz are two candidates to construct the low background interposer because of their good radio-purity; nevertheless, it is non-trivial to produce through silicon vias (TSV) or through quartz vias (TQV) on the large area silicon or quartz wafer. In this work, based on double-sided TSV interconnect technology, we developed the first prototype of a silicon interposer with a size of 10~cm$\times$10~cm and a thickness of 320~$\mu$m. The electrical properties of the interposer are carefully evaluated at room temperature, and its performance is also examined at -110~$^\circ$C with an integrated SiPM on the interposer. The testing results reveal quite promising performance of the prototype, and the single photoelectron signals can be clearly observed from the SiPM. The features of the observed signals are comparable with those from the SiPM mounted on a normal FR4-based PCB. Based on the success of the silicon interposer prototype, we started the follow-up studies that aimed to further improve the performance and yield of the silicon interposer, and eventually to provide a solution for building an ultralow background photodetector system.
\end{abstract}


\section{Introduction}
\label{sec:introduction}
\IEEEPARstart{U}{ltralow} radioactivity backgrounds are a stringent requirement in projects that aim to search for rare fundamental physics phenomena, such as double beta decay, dark matter, solar neutrinos and other events with long half-lives. It is quite challenging to build a detector with an ultralow radioactivity background because it requires a large amount of effort to carefully conduct material screening and to develop radioassay approaches for samples with ultralow contamination. A deep overburden is also required to shield the cosmic rays and reduce muon-induced radioactive isotopes in the detectors. A background-free measurement is an ultimate goal of low background experiments. In recent decades, some low-radioactivity background techniques have been well developed and used to construct low background detectors~\cite{Heusser}\cite{10.3389/fphy.2020.577734}, in which the noble liquid TPC is a competitive technology compared with others in the searches for double beta decay and dark matter~\cite{Chepel:2012sj,MarrodanUndagoitia:2015veg,Aprile:2009dv} and is also proposed in the next generation of multi-ton-scale experiments~\cite{nEXO:2018ylp,darkside,darwin,Avasthi:2021lgy}. With the TPC, information on the scintillation photons and the ionization charge is essential to reconstruct the interaction vertex and energy of particles of interest. Therefore, a photodetector system is an indispensable system in TPCs to collect scintillation photons with sufficient efficiency. In some dual-phase TPCs, the photodetector is also used to detect the photons generated from electroluminescence of the ionization charge.

A typical photodetector module consists of a photosensor, readout electronics and an interposer that provides mechanical support and connections between the sensors and the electronics. A silicon photomultiplier (SiPM)~\cite{sipm} provides an ideal solution for a photosensor that can be produced with an ultralow background. The SiPM combines the advantages of an avalanche photodiode (APD, good radio-purity) and photomultiplier tube (PMT, capability of single photon detection and high gain). The radio-purity of the bare SiPMs can reach the level of sub parts per trillion (ppt), as reported in~\cite{nEXO:2021ujk}. On the other hand, application-specific integrated circuits (ASICs) also provide a good solution for readout electronics with a background level similar to that of SiPM~\cite{nEXO:2021ujk} because of the usage of materials with good radio-purity and good control of the contamination during fabrication. However, it remains an open issue to develop a radio-pure interposer. Common printed circuit boards (PCBs) based on FR4, ceramic, etc., materials are not applicable for this purpose due to their high radioactivity background at a level of parts per million (ppm). The DarkSide experiment uses the interposer made of Arlon 55NT with an intrinsic radioactivity at the level of tens of parts per billion (ppb)~\cite{Kochanek:2020hmq}. However, the radio-purity of Arlon 55NT cannot meet the background requirements in some applications, such as double beta decay searches, which typically require backgrounds to be at the level of ppt or even lower. Quartz and silicon are two ideal materials to build low background interposers due to their high purity. The silicon interposer is more preferable because of exhibiting the same coefficient of thermal expansion (CTE) as the SiPMs, which leads to a lower assembly risk from the mechanical point of view~\cite{murayama2013warpage}, particularly during operation at noble liquid temperatures. The through silicon via (TSV) is essential on the silicon interposer to enhance the coverage of photosensors. The TSVs connect the SiPM sensors on one side to the readout chip on the other side of the interposer. Meanwhile, an interposer with sufficient area and thickness is also required to guarantee fewer interconnections and sufficient mechanical strength. However, it is nontrivial to produce TSVs in a large area interposer with good qualities, and the relevant technologies are not well defined. In this work, we report the development of the first prototype of the silicon interposer with dimensions of $10\times10$~$cm^2$, fabricated by the Institute of Microelectronics (IME) and Institute of High Energy Physics in China. The prototype is evaluated both at room temperatures and liquid xenon temperatures. It shows good electrical performance, and the signals of a single photoelectron can be clearly observed from the integrated SiPMs on the interposer. The promising testing results provide a new solution to build a photodetector system with an ultralow radioactivity background.

This paper is organized as follows: a conceptual design of a photodetector module with ultralow background is first introduced, followed by a description of the silicon interposer design and the material composition of the interposer prototype. The detailed processes of interposer fabrication are then presented. Finally, the prototype silicon interposer is characterized, and the testing results are reported.

\section{Concept of an ultralow background photodetector module}
\begin{figure}
\centering
\includegraphics[width=3.5in]{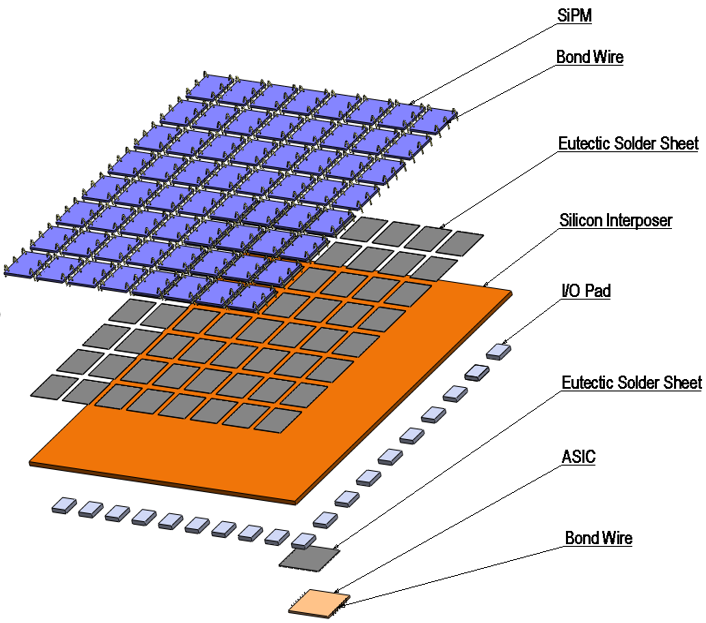}
\caption{Conceptual layout design of a photodetector module with an ultralow radioactivity background.}
\label{pd_integ}
\end{figure}

Fig.~\ref{pd_integ} shows a conceptual layout design of the ultralow background photodetector module. The dimensions of the interposer are designed to be $10~cm\times10~cm$ with a thickness of $\sim$300~$\mu$m. A SiPM array, consisting of $8\times8$ devices (assuming $10\times10$~$mm^2$ per device), is soldered on the front side of the interposer via the eutectic solder sheet and wire bonding for the case of SiPMs without TSVs. SiPMs with TSVs can be soldered via the reflow soldering process. Then, the anode and the cathode of each individual SiPM or a SiPM group (after ganging) are connected to the input-output (I/O) pads on the back side of the interposer via TSVs and signal traces. I/O pads are connected to the corresponding pads on the ASIC chip via wire bonding. The ASIC chip could be directly soldered on the back of the interposer or soldered on an interface board. For the former case, multiple  redistribution layers (RDLs) are required on the back of the interposer to provide sufficient shielding and eliminate the cross talk between the interposer and the chip. For the latter case, the number of RDLs is much more flexible since the RDLs can be produced on the interface boards. The interface board could be fixed on the back of the interposer by means of soldering or other proper technologies. Eventually, the signals from the output of the ASIC chip are sent out to the following readout system through low background cables, such as polyimide flexible cables, which have been proven to have good radio-purity~\cite{kapton}.

\section{Interposer design}
\subsection{Layout design}
\begin{figure}
\centering
\includegraphics[width=3.5in]{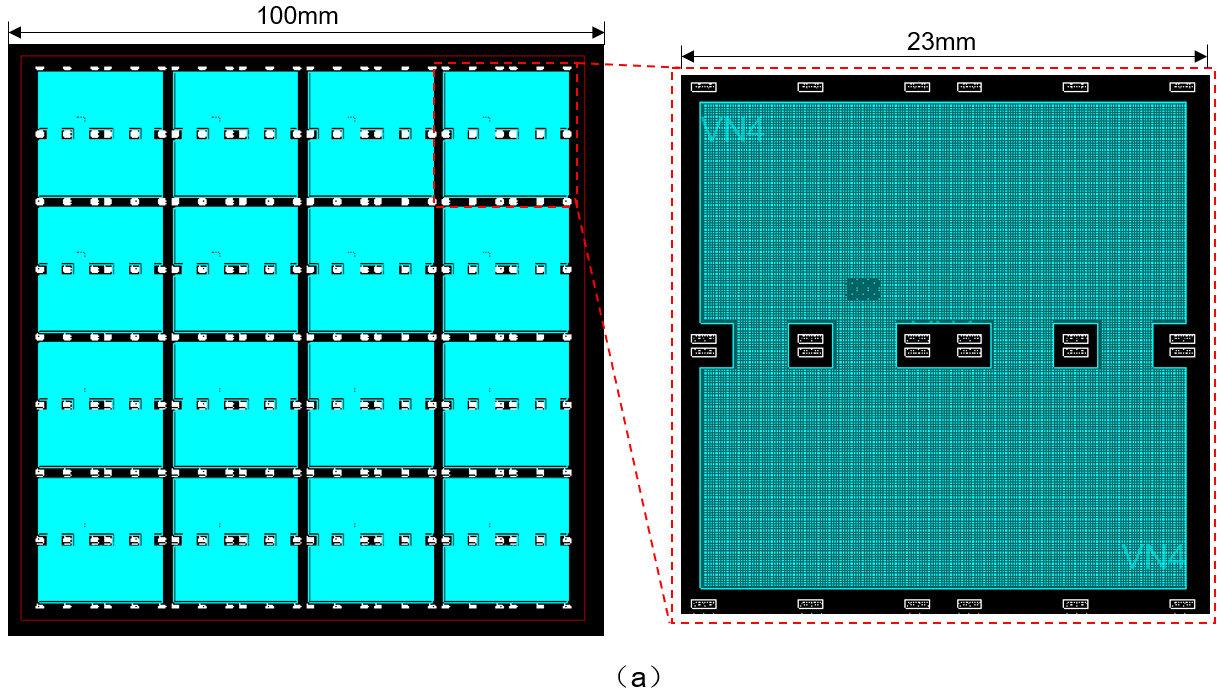}
\includegraphics[width=3.5in]{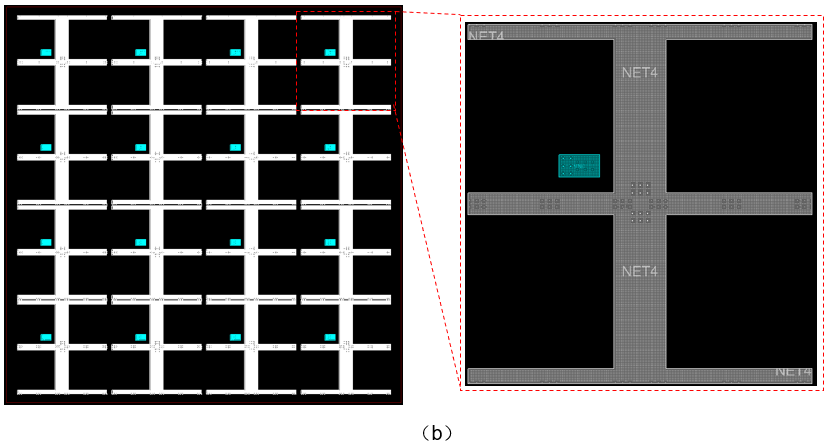}
\includegraphics[width=3.5in]{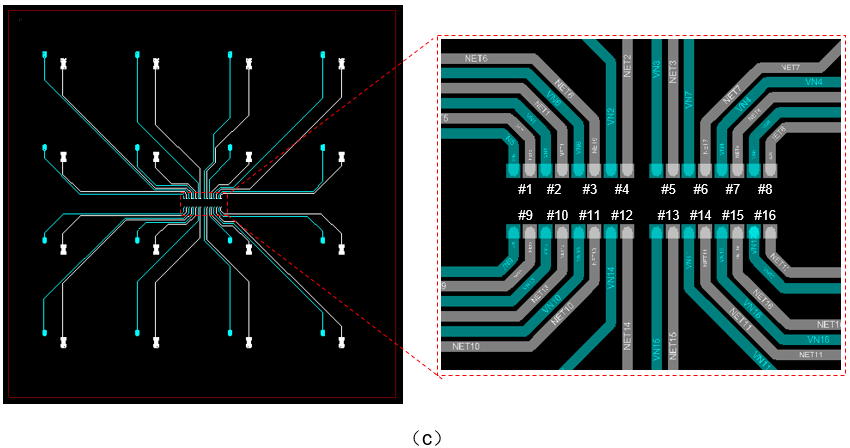}
\caption{The layout design of the silicon interposer prototype, including the first layer RDL on the front (a), the first layer RDL on the back (b) and the second layer RDL on the back (c).}
\label{ly_design}
\end{figure}

A finer line spacing and I/O pitch can be achieved on the silicon interposer~\cite{usman2017interposer} compared with the traditional PCBs, which provides more flexibility for the layout design. In common cases, the maximum area of the silicon interposer used in industry is at the level of a few hundred square millimeters, and the thickness is generally less than 100~$\mu$m~\cite{banijamali2011advanced,lee2016overview,khan2008development}. In this work, the prototype of a silicon interposer is designed to be 100~mm$\times$100~mm with a thickness of 320~$\mu$m. Fig.~\ref{ly_design} (a) shows the layout design of the front side of the interposer, together with the magnified structure. The front side is used to assemble SiPMs, which consist of 16 blocks, represented by the areas filled with the cyan color. Each block is 23~mm $\times$ 23~mm, on which 4 SiPMs can be assembled, assuming SiPM dimensions of 10~mm $\times$ 10~mm, such as the one produced by FBK for the nEXO project~\cite{ako_talk}. More detailed structures in each block are shown in the magnified image. The area filled with the cyan color represents the copper sheet, which means that 4 SiPMs are connected in parallel. The small rectangle with dark cyan color includes 6 TSVs that connect the copper sheet to the back side of the interposer. This redundant design not only improves the reliability of the TSV but also reduces the resistance on the signal transmission line induced by TSVs. Six pads are designed with a size of 200~$\mu$m $\times$ 400~$\mu$m to connect the anode on the front side of one SiPM via wire bonding. Therefore, there are 18 pads in total in each block. Each pad consists of 3 TSVs.

The layout design on the back of the interposer is shown in Fig.~\ref{ly_design} (b) and Fig.~\ref{ly_design} (c), which display the first and second redistribution layers (RDLs) on the back. The first RDL also includes 16 blocks that are connected with the corresponding 16 blocks on the front side through TSVs. The zoomed in image is the first RDL in one block on the back of the interposer. The area filled with the cyan color represents the pad that is connected to the copper sheet on the front side, and the area filled with the gray color represents the copper traces that are connected to the 24 pads on the front side. The 24 pads, corresponding to 4 SiPMs in one block, converge in this layer. In the second RDL, 16 pairs of pads from the 16 blocks are connected to the 32 I/O pads located at the center of the interposer via the metal traces. The width of the metal trace is 400~$\mu$m, and the thickness is 5~$\mu$m. The lengths of traces are not equal for different connections; however, they can be designed to have equal lengths if precise timing measurement is required in applications. The detailed structure of the I/O pads is shown in the magnified figure. The I/O pads (the cyan pad is connected to the anode of the SiPM, and the gray pad is connected to the cathode of the SiPM) with dimensions of 400~$\mu$m $\times$ 400~$\mu$m are arranged in two rows and are used to connect to the readout electronics.

\subsection{Structure design}
\begin{figure}
\centering
\includegraphics[width=3.5in]{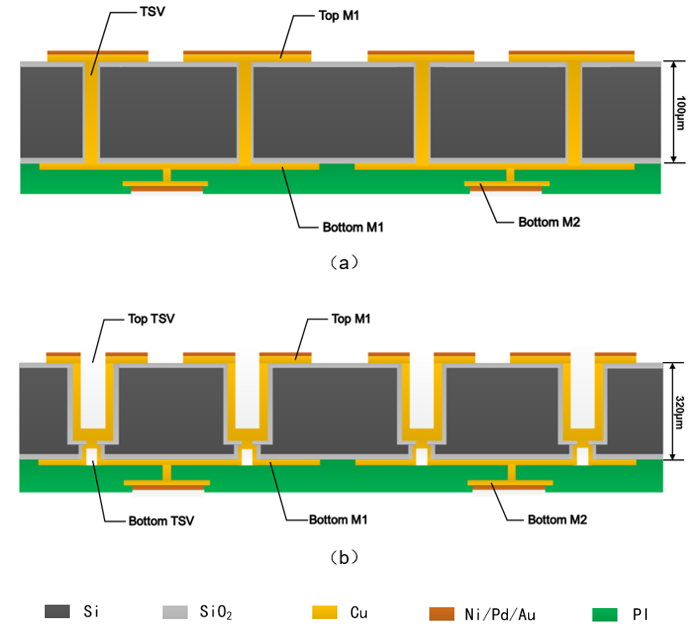}
\caption{(a) The structure design of a typical silicon interposer. (b) The new structure design proposed in this work based on double-sided TSV interconnection technology. The materials are indicated with different colors shown at the bottom.}
\label{st_design}
\end{figure}

TSV is the most critical process in silicon interposer fabrication~\cite{velenis2009impact}, which is also an important factor in the considerations of structure design. A typical structure design of an interposer is shown in Fig.~\ref{st_design} (a). This design is suitable with thin and small interposers; however, for a large and thick interposer, such as the one proposed in this work, it is quite challenging to use this common structure because the TSV formation becomes difficult. In general, the TSV formation involves four steps, including TSV etching, deposition of the dielectric layer, TSV metallization, and TSV revealing on the back side. To produce the common TSV structure on the large interposer, the following two issues are foreseen.

In terms of TSV etching and metallization, a deeper TSV requires a larger diameter of the TSV due to the limit of the aspect ratio as determined by the current technologies, with a typical value of 10. However, it is more risky to produce TSVs with larger diameters because the failure rate increases and the electroplating efficiency decreases. Meanwhile, it is also challenging to obtain fully filled TSVs with a depth of more than 300~$\mu$m.

In terms of TSV revealing, the most critical step is to use the chemical-mechanical polishing (CMP) process to grind and polish the dielectric and metal on the back side of the interposer. This process requires the thickness uniformity of the remaining silicon after dry etching to be controlled within 2~$\mu$m~\cite{crook2013dielectric,huang2013integration,kumar2012robust}; however, it is difficult to meet such requirements for deep TSVs at levels of 300~$\mu$m in the entire wafer.

Because of the two aforementioned difficulties, we decided to discard the technological route used for the common cases. Meanwhile, we propose a new structure design based on double-sided TSV interconnection technology, as shown in Fig.~\ref{st_design} (b). In the new design, the front side and back side of the interposer have two types of TSVs with different dimensions. The top TSV (T-TSV) has a diameter of 80$\mu$m and a depth of 300$\mu$m (80×300$\mu$m), while the bottom TSV (B-TSV) has a diameter of 50$\mu$m and a depth of 20$\mu$m (50×20$\mu$m). The T-TSV and B-TSV are interconnected inside the silicon substrate. Silicon dioxide is used as the insulation layer to isolate the metal in the TSVs from the silicon substrate. Two RDLs are designed on the back of the interposer. Again, silicon dioxide is introduced to separate the first RDL from the silicon substrate. The polyimide (PI) provides insulation between the first layer and the second layer of RDLs that establish connections through vias. The relevant technologies are defined and developed to complete the interposer fabrication.

\subsection{Material composition and radio-purity}
The radioactivity of the interposer is determined by the radio-purity of the raw materials and the contamination from the environment and equipment during fabrication. The latter can be generally well controlled, since all production processes are conducted in class 1000 clean rooms or even better. Regarding the raw materials, the interposer is composed of silicon, silicon dioxide, PI, copper, titanium, nickel, palladium, and gold. Based on the layout design and the structure design, the mass fraction of each material for a single interposer is estimated and summarized in Table~\ref{mat_com}. More than 90\% of the mass is contributed by the silicon substrate, and the copper used for tracing contributes approximately 5.7\%. The nickel used in the bonding pads comes third with $\sim$2\%. The mass fractions of other materials are less than 1\%.

The radio-purity of the fabricated interposer prototype was measured by inductively coupled plasma mass spectrometry (ICP-MS) at Pacific Northwest National Laboratory (PNNL). Due to the composite nature of the interposer devices the assay involved a stepwise approach to digest the variety of materials in the device to better understand the radio-purity of the different materials outlined above. In order to mitigate any contamination from subsampling the material for assay, the interposer device was cut into $\sim$2~cm square pieces with a laser engraver. Some of the pieces were then rinsed with DI water 3 times and allowed to dry overnight. Both washed and unwashed pieces were subsampled into validated perfluoryl alkoxy (PFA) Savillex (Eden Prairie, MN, USA) vials to be assayed in a stepwise manner to ascertain levels of impurities in a piece meal approach that grouped different components into three categories: metals fraction, silicon/silica/titanium fraction, and the polyimide fraction.  
For the metals fraction, a known amount of non-natural $^{229}$Th and $^{233}$U tracer was spiked into each sample and process blank. Samples were etched in Optima grade aqua regia, and the solution was transferred into separate vials and dried down and re-suspended in 2\% Optima grade nitric acid. After a 10X dilution, the samples were analysed with an Agilent 8900 ICP-MS using an O$_{2}$ reaction gas to avoid the polyatomic interferences present in the sample, following previous work by our PNNL colleagues~\cite{D0JA00220H}.
Quantitation of $^{232}$Th and $^{238}$U was performed using isotope dilution mass spectrometric methods.
For the silicon/silica/titanium fraction, another addition of a known amount of non-natural $^{229}$Th and $^{233}$U tracer was spiked into the remaining sample and process blanks. The remaining soluble materials (Si, SiO$_{2}$, and Ti) were dissolved in a mixture of Optima grade HF and 8N HNO$_{3}$. The solution was transferred into separate vials and dried down and re-suspended in 2\% Optima grade nitric acid. The samples were then analysed with an Agilent 8800 ICP-MS and quantitation of $^{232}$Th and $^{238}$U was performed using isotope dilution mass spectrometric methods.
Finally, the remaining PI was placed into microwave vessels and a known amount of non-natural $^{229}$Th and $^{233}$U tracer was spiked into the remaining sample and process blanks. The samples were digested at 250°C in full strength Optima grade HNO$_{3}$ using a CEM Mars 6 microwave digestion system (Matthews, NC, USA), following a similar approach by our PNNL colleagues in Ref.~\cite{ARNQUIST2020163573}.
Samples were then dried down and re-suspended in 2\% HNO$_{3}$ and analyzed on an Agilent 8900 ICP-MS. Quantitation of $^{232}$Th and $^{238}$U was performed using isotope dilution mass spectrometric methods. 
The measured contents of $^{238}$U and $^{232}$Th were found to be 4.6~ppt and 2.3~ppt for the overall interposer, respectively. There was no significant difference between washed and unwashed subsamples showing the interposer processing and handling steps, as well as the assay subsampling methods, were extremely clean and provided negligible contamination overall. The metals fraction looked to be the dominant driver for the contamination in the low single digit ppt range, while the silicon/silica/titanium and PI fractions were determined with upper limits $<$1-2~ppt for both $^{238}$U and $^{232}$Th. We are quite pleased with these results, as radio-purity targets in the low ppt range are desired, and reaching this level with a complex device is a significant achievement. The radioassay of raw materials will also be carried out in the near future. 

\begin{table}
\centering
\caption{Material composition of the silicon interposer prototype}
\label{mat_com}
\setlength{\tabcolsep}{6pt}
\begin{tabular}{|c|c|c|}
\hline
Materials & Total mass (mg) & Mass fraction (\%) \\
\hline
Silicon & 7448.2 & 90.45 \\
Copper & 471.6  & 5.73 \\
Nickel & 177.6 & 2.16 \\
Polyamide & 177.6 & 0.84 \\
Silicon dioxide & 25.4 & 0.31 \\
Palladium & 24 & 0.29 \\
Titanium & 15 & 0.18 \\
Gold & 3.9 & 0.05 \\
\hline
\end{tabular}
\end{table}

\section{Interposer fabrication}
\begin{figure}
\centering
\includegraphics[width=3.5in]{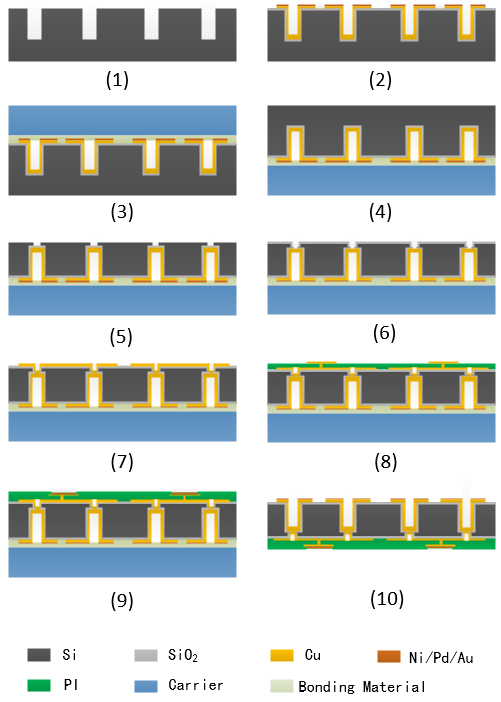}
\caption{The process flow of interposer fabrication based on double-sided TSV interconnection technology. The interposer materials are represented as different colors shown at the bottom.}
\label{fab_process}
\end{figure}

The fabrication processes, based on double-sided TSV interconnection technology, are investigated and defined in this work. In total, 10 processes are involved during the fabrication. They are shown in Fig.~\ref{fab_process}, and the materials are represented as different colors labeled at the bottom of the figure. The technical details during the fabrication will be reported in a dedicated paper and the fabrication processes are briefly summarized as follows:

\begin{enumerate}
\item T-TSV photolithography and deep reactive ion etching (DRIE) by the Bosch process. The Bosch process can realize high selectivity, a high etch rate and high aspect ratio dry etching. The etching parameters and passivation steps are optimized to minimize the sidewall roughness and improve the depth uniformity of TSVs. The smooth sidewall with no curvature effect and no undercut guarantees success of the subsequent processes of passivation layer deposition and metallization.
\item A layer of SiO$_2$ that serves as an insulation layer is first deposited after T-TSV etching by plasma-enhanced chemical vapor deposition (PECVD). Then, T-TSV metallization and RDL are performed with conformal electroplating. Eventually, electroless plating is used to form a layer of nickel-palladium-gold for metal finishing.
\item Temporary bonding. When the front side process of the wafer is completed, a carrier wafer is bonded to the front side of the wafer by the temporary bonding process.
\item Back side wafer grinding. The wafer is thinned to a proper thickness by wafer grinding while maintaining planarity.
\item B-TSV photolithography and etching by the Bosch process.
\item PECVD is conducted to deposit a layer of SiO$_2$ as the insulation layer for B-TSVs. However, the SiO$_2$ at the bottom of the B-TSV is removed by dry etching, together with the SiO$_2$ at the bottom of the T-TSV. Then, a window is opened and made ready for conducting the connections between the B-TSVs on one side and the T-TSVs on the other side.
\item B-TSVs are interconnected with the T-TSVs by conformal plating, while the first RDL (Bottom M1) on the back is also completed.
\item A layer of PI is formed by photolithography on the top of the first RDL as the insulating layer, and then electroplating is performed to form the second RDL.
\item Electroless plating and solder mask coating. The second RDL is protected by the solder mask, except for the I/O pads.
\item Temporary bonding removal and wafer dicing.
\end{enumerate}

\begin{figure}
\centering
\includegraphics[width=3.5in]{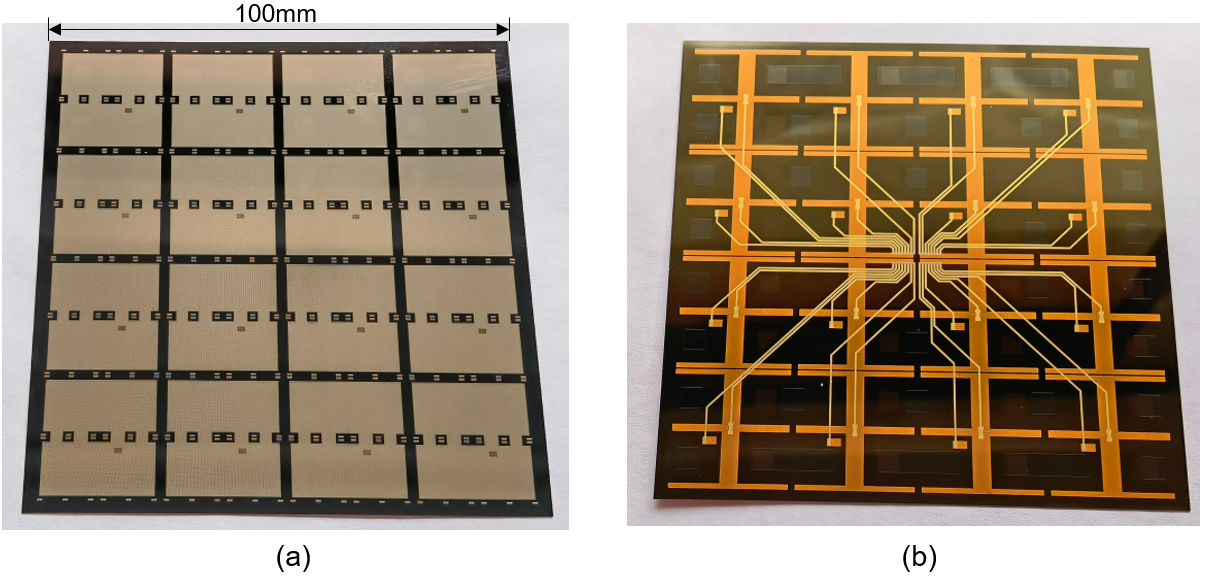}
\caption{The front side (a) and back side (b) of the interposer prototype fabricated based on double-sided TSV interconnection technology with dimensions of 100~mm $\times$ 100~mm.}
\label{pro_pic}
\end{figure}

\begin{figure}
\centering
\includegraphics[width=3.5in]{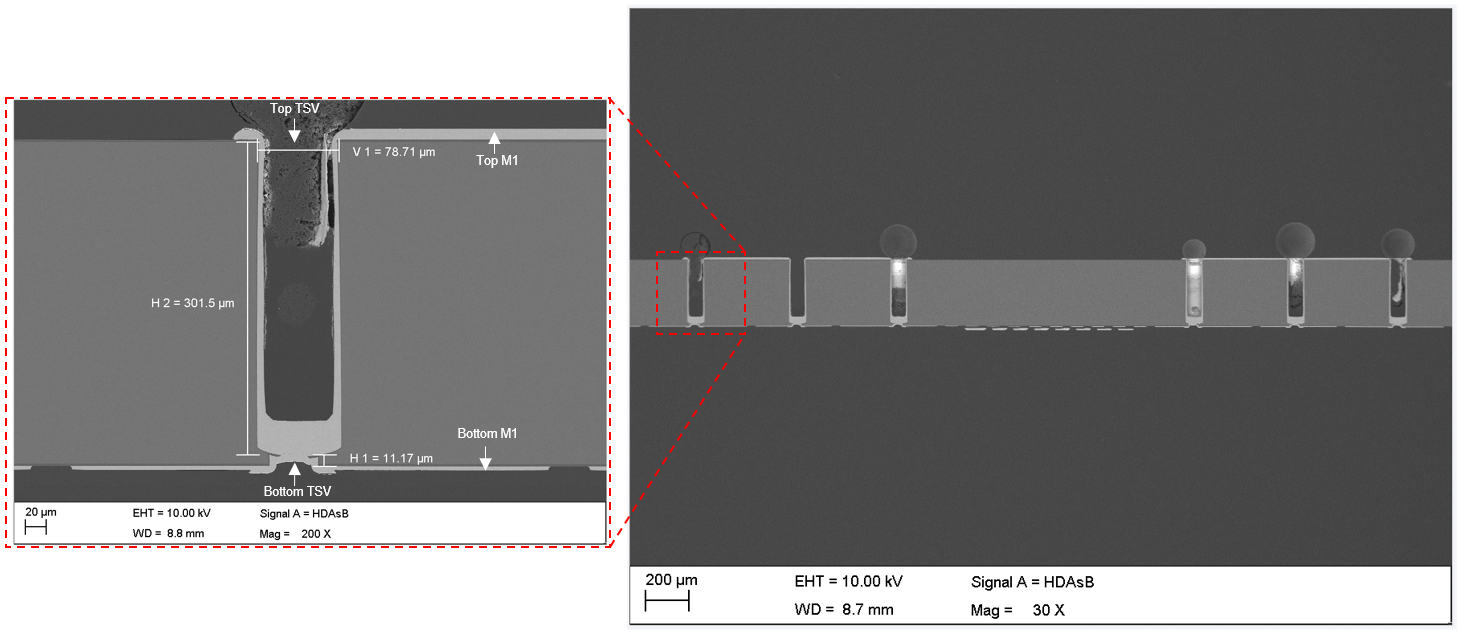}
\caption{The cross section of the interposer prototype inspected via SEM. The enlarged TSV structure is shown on the left.}
\label{interposer_cs}
\end{figure}

With the aforementioned processes, we fabricated the first silicon interposer prototype with dimensions of 100~mm $\times$ 100~mm, which is shown in Fig.~\ref{pro_pic}. Fig.~\ref{pro_pic} (a) shows the front side of the interposer prototype, and Fig.~\ref{pro_pic} (b) shows its back side. No defects are found on the prototype from the visual checks, and the layout is consistent with the design. The cut sections are inspected via scanning electron microscope (SEM), and the cross-section is shown in Fig.~\ref{interposer_cs}, together with the enlarged TSV structure. The total thickness of the interposer is measured to be 312.7~$\mu$m. The depth of the T-TSV is 301.5~$\mu$m, and the diameter is 78.7~$\mu$m. Regarding the B-TSV, the depth and diameter are 11.2~$\mu$m and 49.9~$\mu$m, respectively. The dimensions of the TSVs are consistent with the design, except for the depth of the B-TSVs, which is only half of the designed value. This discrepancy is caused by the imperfect control in the wafer thinning process. However, we do not expect the discrepancy to significantly affect the mechanical and electrical performance of the interposer. From the enlarged TSV structure, we found that the T-TSV and B-TSV are successfully connected.

\section{Electrical performance}
The leakage currents of blocks and the resistance of the signal transmission lines are the two critical electrical specifications of the interposer, which are required to be well controlled. The leakage current can induce a voltage drop on the filter resistor in the circuit of SiPMs' bias voltage input, so it alters the bias voltage on SiPMs and results in gain variations, which can worsen the resolution of single photoelectron detection. The resistance on the signal transmission line could significantly affect the electronic noise. In general, the insulation performance of a good interposer should be better than several hundred gigaohms, and the resistance on traces should be less than a few ohms. In some applications, series connections are used in SiPM ganging to reduce input capacitance. Therefore, the interposer is required to afford a much higher bias voltage and guarantee that no breakdown may occur.

\subsection{Leakage current test}
\begin{figure}
\centering
\includegraphics[width=3.5in]{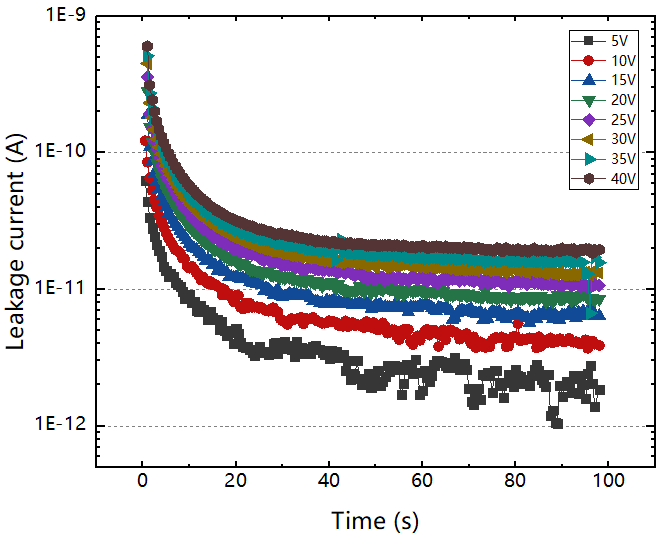}
\caption{Leakage currents as a function of time measured for block \#12 with different DC voltages marked with different colors and markers.}
\label{leakage_current_time}
\end{figure}

\begin{figure}
\centering
\includegraphics[width=3.5in]{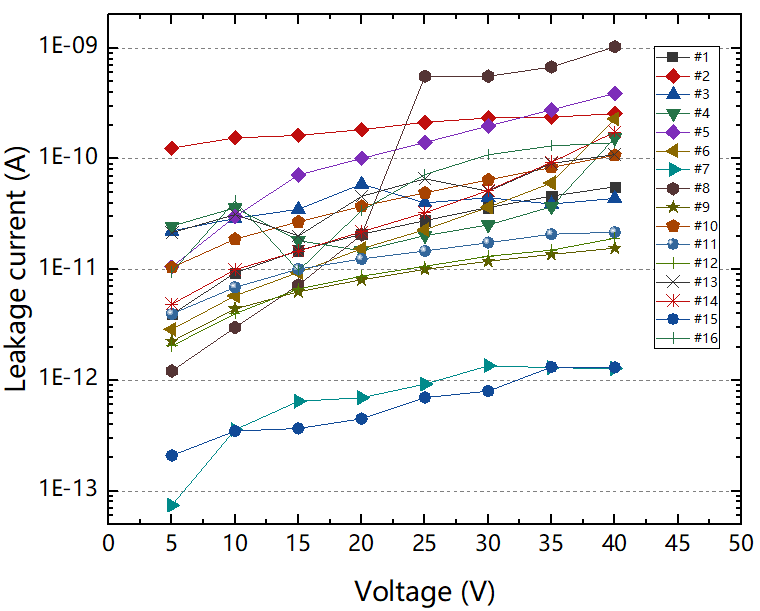}
\caption{Leakage currents of all 16 blocks on the interposer as a function of DC voltage.}
\label{leakage_current_voltage}
\end{figure}

The leakage current is an important figure of merit to verify the quality of the insulation layer and the maturity of the TSV processes. The leakage currents of 16 independent blocks on the interposer were measured by using a probe station and a Keithley 6487 picoammeter. No SiPM devices were integrated on the interposer during measurement. The direct-current (DC) voltage, provided by the picoammeter, was applied to the two I/O pads of each block located on the back side of the interposer. The two pads are designed to connect the anode and cathode of SiPMs. A typical leakage current as a function of time is shown in Fig.~\ref{leakage_current_time}, which is measured for block \#12 with different DC voltages marked with different colors and markers, ranging from 5~V to 40~V. The measured leakage currents gradually decrease over time and become stable after tens of seconds. This feature is caused by the settling time of the circuit, which is affected by the shunt capacitance induced by the connecting cables, test fixtures, and the interposer under test. The data points in the flat region of the curves are averaged to obtain the final leakage current of the tested block. Fig.~\ref{leakage_current_voltage} shows the leakage currents as a function of DC voltage conducted for all 16 blocks. The leakage currents of all blocks are less than 1~nA at the maximum measured voltage of 40~V. From this information, the resistances are calculated and listed in Table~\ref{insul_r} for all 16 blocks. The insulation resistances are centralized at approximately 500~G$\Omega$, and the minimal resistance is 125~G$\Omega$, which appears in block \#2.

\begin{table}
\centering
\caption{Insulation resistances of different blocks on the interposer.}
\label{insul_r}
\setlength{\tabcolsep}{6pt}
\begin{tabular}{|c|c|c|c|}
\hline
Block ID & R (G$\Omega$) & Block ID & R (G$\Omega$) \\
\hline
1 & 861 & 9 & 2500 \\
2 & 125  & 10 & 475 \\
3 & 644 & 11 & 1670 \\
4 & 965 & 12 & 2260 \\
5 & 161 & 13 & 482 \\
6 & 924 & 14 & 647 \\
7 & 26500 & 15 & 35900 \\
8 & 537 & 16 & 548 \\
\hline
\end{tabular}
\end{table}

A few interposers with different configurations are fabricated to investigate the reasons for the relatively low insulation resistance. Some of them include only the RDL, together with T-TSVs, on the front side and no structures on the back side. With this configuration, the insulation performance of the silicon dioxide on the front side could be examined. In contrast, others contain only one or two RDLs with B-TSVs on the back side and no structures on the front side, so the insulation of the silicon dioxide and PI on the back side could be verified. After examinations, we found that both silicon dioxide and PI achieve excellent insulation performance larger than T$\Omega$. Therefore, we conclude that the connection regions between the T-TSVs and B-TSVs cause a small leakage current, which could be further improved in future research.

\begin{figure}
\centering
\includegraphics[width=3.5in]{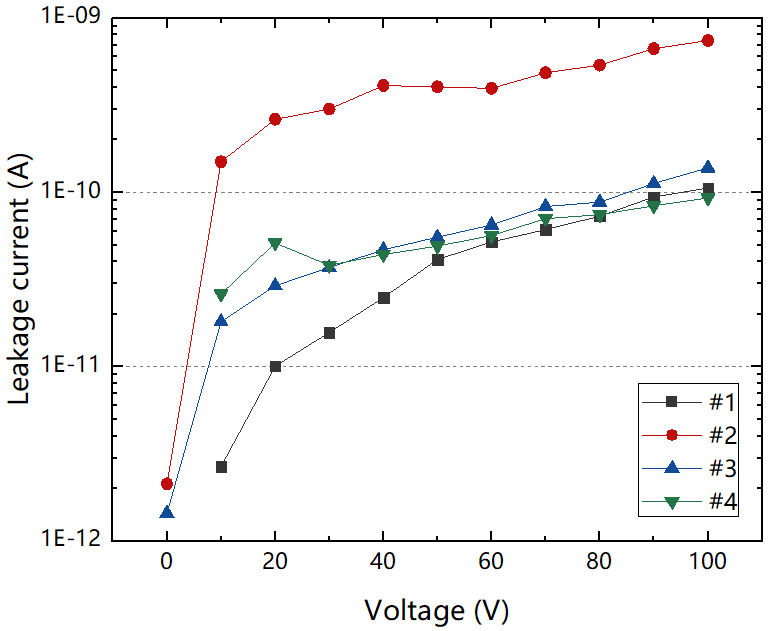}
\caption{The leakage currents as a function of DC voltage detected for blocks \#1 (black), \#2 (red), \#3 (blue), and \#4 (cyan).}
\label{breakdown_test}
\end{figure}

Four blocks (\#1, \#2, \#3, and \#4) are selected and applied with much higher DC voltages. The leakage currents as a function of voltage are shown in Fig.~\ref{breakdown_test}. The breakdown phenomenon is not observed for the 4 tested blocks with voltages of up to 100~V. Since the smallest thickness of the insulation layer is $\sim$400~nm in TSVs, the interposer can afford an electric field of at least 2.5$\times$10$^{6}$~V/cm without any breakdowns. A destructive test will also be performed to determine the breakdown voltage after completing the interposer's other missions.

\subsection{Continuity test}
\begin{figure}
\centering
\includegraphics[width=3.5in]{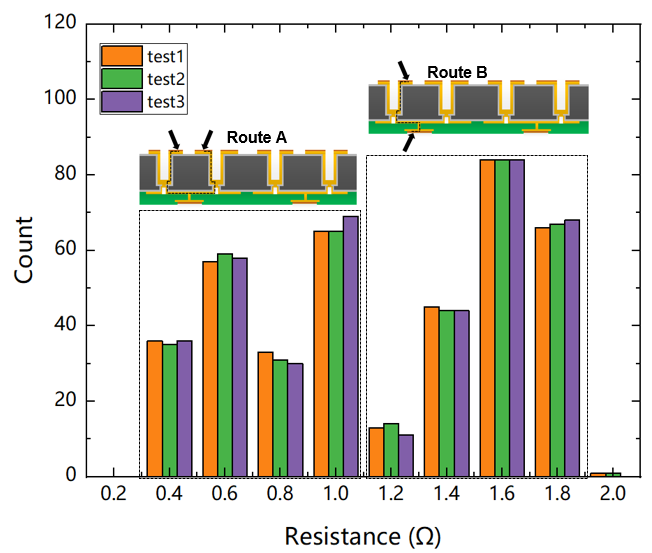}
\caption{Resistance of route A and route B in 16 blocks. The measurements are repeated 3 times and are represented as different colors.}
\label{r_trace}
\end{figure}

A flying probe machine produced by ATG was used to measure the resistance of the two different routes (A and B) on the interposer. The probe connections of the two routes are shown in Fig.~\ref{r_trace}. The black arrows indicate the probes. Route A includes the full signal transmission lines, in which the vias in PI and the second RDL on the back side of the interposer are not included. However, route B includes all signal transmission lines. The resistances of routes A and B are shown in Fig.~\ref{r_trace}, distinguished by the dashed boxes. These measurements are performed 3 times and are indicated as different colors in the figure. No disconnections are found in the interposer. The maximum resistance is less than 2~$\Omega$ on the interposer. The resistance of route A is approximately half of the resistance of route B because of its shorter traces ($\sim$3.4~cm on average) compared with those ($\sim$7.3~cm on average) of route B. With inputs of the dimension information of traces, the resistance of traces is subtracted from the measured resistance, which yields the resistance induced by the TSVs of $\sim$0.011~$\Omega$. In fact, the resistance can be further reduced by increasing the width and thickness of traces or shortening their lengths.

\section{Performance with integrated SiPMs}
The layout of the prototype interposer is designed for bare SiPMs with dimensions of 10~mm$\times$10~mm that are manufactured by FBK. However, because no proper SiPMs are available in our hands, we integrate a SiPM array produced by Hamamatsu on the interposer with a series number of S13371-6050CQ-02, consisting of 2$\times$2 devices (approximately 6~mm$\times$6~mm each). The array is enclosed in a ceramic jacket with a quartz window, and 8 pins on the back of the array are connected to the anodes and cathodes of the 4 devices. The array is inserted on an interface board, in which the 4 devices are connected in parallel. Then, two short cables connect the cathode and anode on the interface board to the corresponding pads of the \#2 block on the interposer. In the end, another two cables are used to connect the two I/O pads on the back of the interposer to the picoammeter or an amplifier board for I-V measurements and waveform data collection. We also mount another SiPM array from the same production batch on a normal FR4-based PCB, which serves as a reference device for the comparisons. After integration, we first measured the I-V curves for the two SiPMs at different temperatures in the dark. Then, a commercial digitizer (DT5751) from CAEN was used to collect the waveform from the amplifier board to evaluate the features of single photon detection at -110~$^{\circ}$C for the two SiPMs, in which the SiPMs were illuminated by an LED light source driven by a pulse generator.

\subsection{I-V curves}
\begin{figure}
\centering
\includegraphics[width=3.5in]{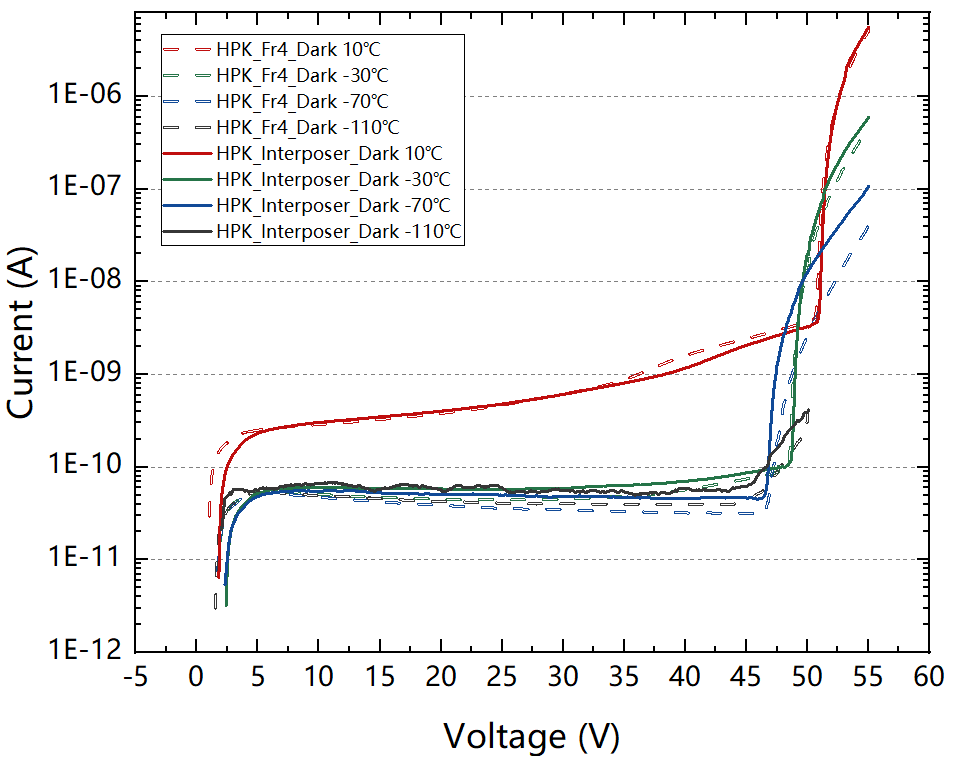}
\caption{I-V curves of the SiPMs on the interposer (solid line) and FR4-based PCB (dotted line) measured at four different temperatures of 10~$^{\circ}$C (red), -30~$^{\circ}$C (green), -70~$^{\circ}$C (blue), and -110~$^{\circ}$C (black). The differences in currents in the region of pre-breakdown are not significant at the low temperatures, because of the aforementioned settling time of the circuit and the imperfect shielding of cables.}
\label{i-v}
\end{figure}

The I-V curves of SiPMs were measured with a Keithley 6487 picoammeter at 4 different temperatures of 10~$^{\circ}$C, -30~$^{\circ}$C, -70~$^{\circ}$C, and -110~$^{\circ}$C. The I-V curves are obtained by performing a fast voltage scan with a time interval of 0.2~s and a voltage step of 0.1~V. The low-temperature environment was provided by a cryogenic box that also served as a dark box. The measured I-V curves in the dark condition are shown in Fig.~\ref{i-v}. The breakdown features, represented by the inflection points of the curves, can be clearly observed for the two tested SiPMs, except the ones measured at -110~$^{\circ}$C, because of SiPMs' extremely low dark count rate. The breakdown voltages decrease when the temperature decreases. This feature agrees with the expectation that has been reported in many other studies. The I-V curves between the SiPMs on the interposer and the FR4 PCB are quite consistent at temperatures above -70~$^{\circ}$C both in the regions of pre- and post-breakdown. In this temperature range, the currents are dominated by the dark noise of SiPMs. At lower temperatures, the visible differences are mainly contributed by the variations in the intrinsic properties of the two SiPMs and by the imperfect light-tight performance of the cryogenic box, in which the latter is dominant. Therefore, from the I-V curves, we found that the performance of the interposer is comparable with that of the FR4-based PCB.

\subsection{Single photoelectron features}
\begin{figure}
\centering
\includegraphics[width=3.5in]{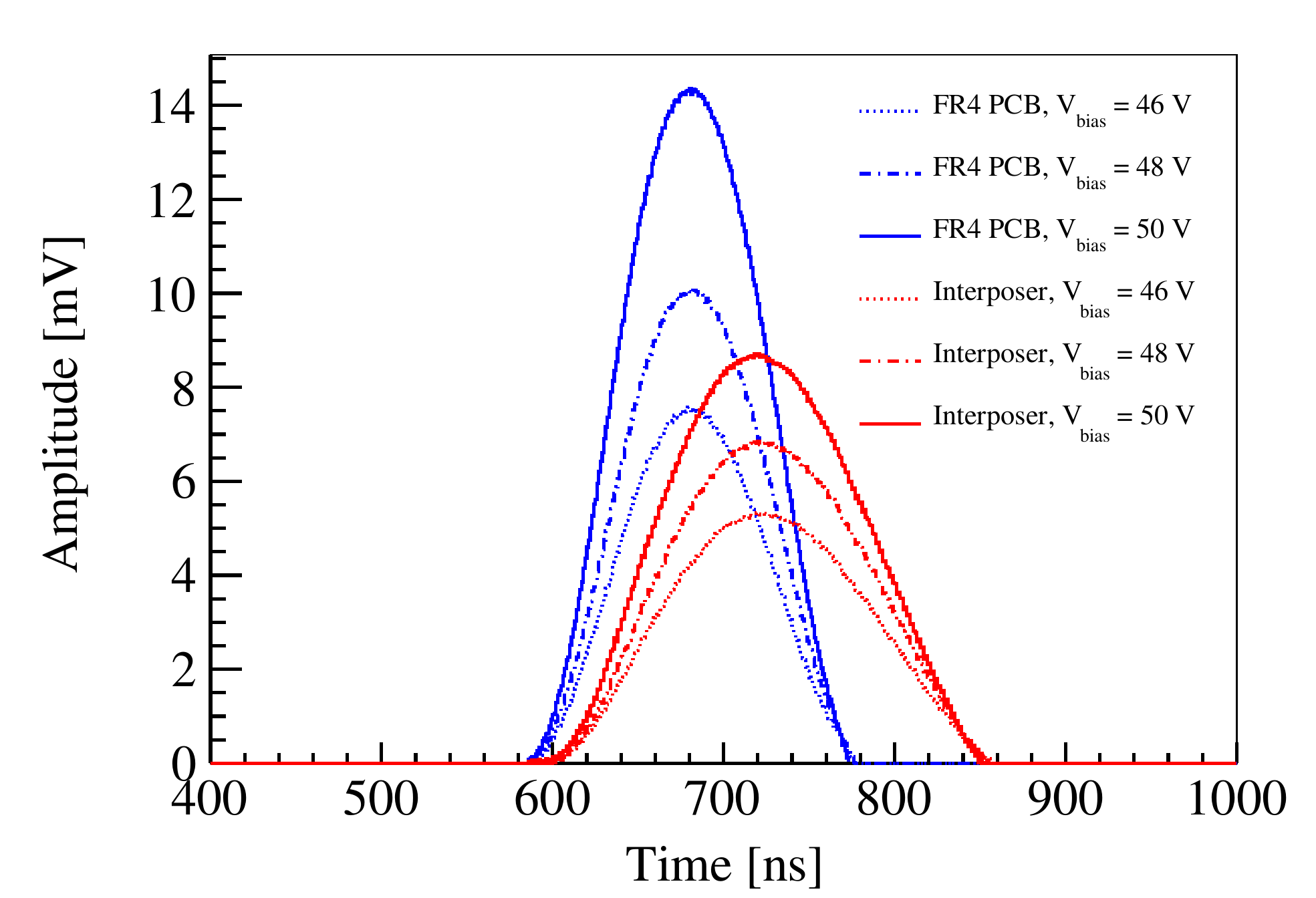}
\caption{Averaged waveform of a single photoelectron for the SiPMs on the interposer (red) and the FR4 PCB (blue) at bias voltages of 46~V, 48~V and 50~V.}
\label{spe_waveform}
\end{figure}

Charges generated during SiPM avalanches are transmitted to the amplifier board via the interposer. The amplified signals are digitized by the CAEN DT5751 with a 1~GHz sampling rate, and the waveform is saved on the disk through data links. The waveform is collected at five different bias voltages of 46~V to 50~V with a step of 1~V at the temperature of -110~$^{\circ}$C. Fig.~\ref{spe_waveform} shows the typical averaged waveform of a single photoelectron for the SiPMs on the interposer (red) and FR4 PCB (blue) at bias voltages of 46~V (dotted line), 48~V (dashed line) and 50~V (solid line). The area (charge) of the waveform at 46~V for the FR4 PCB is normalized to that of the interposer at the same bias voltage; then, the same normalization factor is applied for other voltages for the FR4 PCB. The waveform from the interposer is wider ($\sim$300~ns) than that from the FR4 PCB ($\sim$200~ns). This indicates a larger capacitance at the input end of the amplifier, which might be contributed by the interposer. More detailed studies will be performed to investigate the reasons underlying this observation.

\begin{figure}
\centering
\includegraphics[width=3.5in]{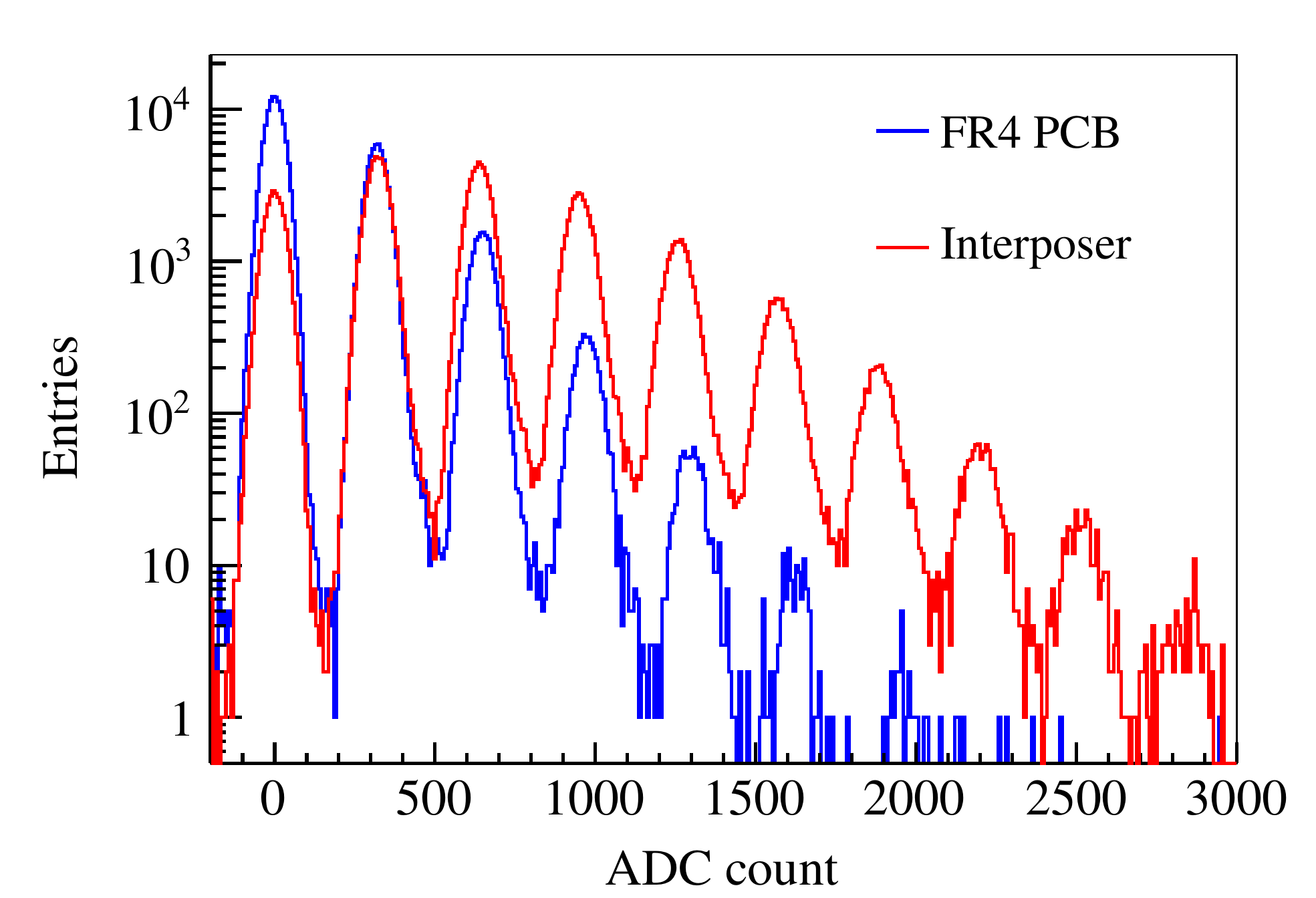}
\caption{Charge spectra of the two SiPMs on the interposer (red) and the FR4 PCB (blue) at a bias voltage of 46~V. The spectrum of the FR4 PCB is scaled up to the gain at the same overvoltage (2.41~V) as that of the interposer, corresponding to a bias voltage of 46.31~V. Different light levels are illuminated on the two SiPMs because the light field in the cryogenic chamber is not uniform.}
\label{Q_spec}
\end{figure}

\begin{figure}
\centering
\includegraphics[width=3.5in]{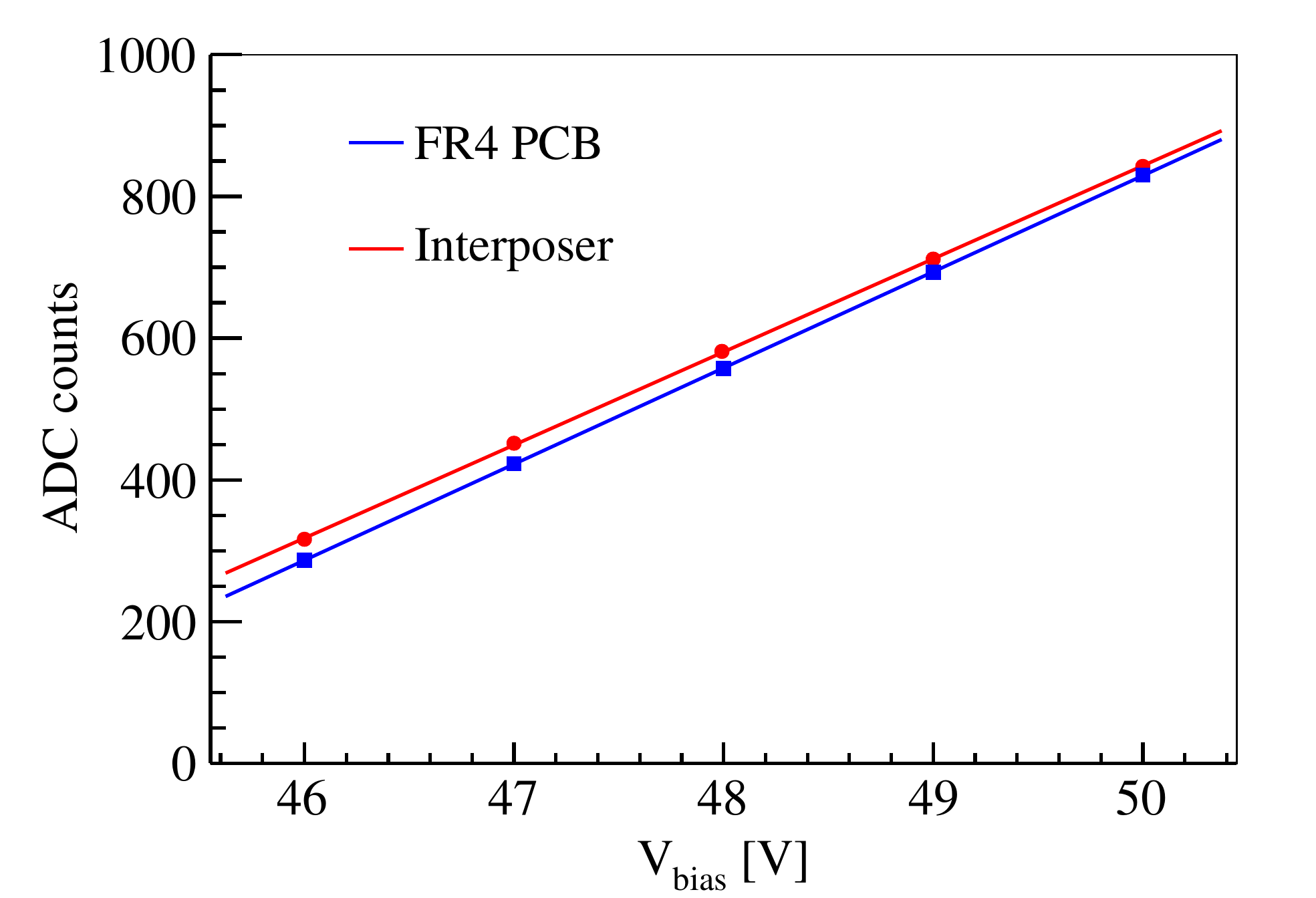}
\caption{Gain as a function of the bias voltage for the two SiPMs on the interposer (red) and the FR4 PCB (blue). The differences are caused by the different breakdown voltages of the two SiPMs, even though they are from the same production batch.}
\label{gain_bias}
\end{figure}

The charge spectra are obtained by performing a simple waveform integration with a time window of 1~$\mu$s. The results are shown in Fig.~\ref{Q_spec} for the SiPMs both on the interposer (red) and the FR4 PCB (blue) at a bias voltage of 46~V. However, because the breakdown voltage of the SiPM on the FR4 PCB is 0.31~V higher than that on the interposer, the spectrum of the FR4 PCB is scaled up to the gain at the same overvoltage as that of the interposer, corresponding to a bias voltage of 46.31~V. The first peak in each spectrum results from the pedestal. The following peaks correspond to one, two, three, or more photoelectrons. We found that the number of detected photons can be effectively identified and distinguished for the two SiPMs, even at the lowest applied bias voltage (2.41~V overvoltage). A better signal-to-noise ratio can be achieved for higher bias voltages. The positions of peaks with the same photoelectron numbers match well between the SiPMs on the interposer and the FR4 PCB. However, more photons are collected by the SiPM on the interposer compared with that on the FR4 PCB. This is caused by the nonuniform distribution of the light field inside the dark box. The breakdown voltage of each SiPM is extracted from Fig.~\ref{gain_bias} by performing a linear fit, which shows the ADC counts of a single photoelectron as a function of bias voltage, together with the fitted lines. The breakdown voltages are determined to be 43.58~V and 43.89~V for the SiPMs on the interposer and FR4 PCB, respectively.

\section{Conclusion and outlook}
It is of great importance to develop a photodetector system with an ultralow radioactivity background in rare event searches, such as projects that aim to search for double beta decays and dark matter based on the noble liquid TPC. The interposer, used to provide support and connections between the photosensors and the electronics, is a bottleneck to build such a radio-pure photodetector. Silicon is an ideal material from which to construct the interposer because of its good purity and good CTE match with silicon-based photosensors, such as SiPMs. However, little information can be found regarding silicon interposer fabrication at the level of hundreds of square centimeters. In this work, based on double-sided interconnected TSV technology, we developed the first prototype of a silicon interposer with dimensions of 100~mm$\times$100~mm$\times$0.32~mm. Detailed characterizations were performed to evaluate the electrical performance of the interposer at room temperature. No connection failures are found in the prototype, and the resistance of all traces, including TSVs, not surpasses 2~$\Omega$. The insulation performance is centralized at approximately 500~G$\Omega$, and the worst insulation performance is 125~G$\Omega$ among the tests of all 16 blocks. The performance with an integrated SiPM array from Hamamatsu is also measured and compared with another SiPM array from the same batch mounted on a normal FR4 PCB. The I-V curves are measured from room temperature down to -110$^{\circ}$C (around the liquid xenon temperature). No significant differences are observed from I-V curves between the interposer and the FR4 PCB. The features of single photoelectron detection are also investigated at -110$^{\circ}$C. The signals of a single photoelectron can be clearly observed and show excellent photon counting capability with the interposer at an overvoltage of 2.41~V; nevertheless, a wider waveform of a single photoelectron is found compared with that from the FR4 PCB. This might be caused by the larger capacitance input to the amplifier from the interposer, which will be further investigated.

Based on the success of the first silicon interposer prototype, we started the ensuing R\&D plans and developments aimed to further improve the insulation quality and uniformity among blocks. The performance of the interposer with fully integrated SiPMs will be carefully evaluated, and more results will be reported in the near future.

\section*{Acknowledgments}
We gratefully acknowledge support from the National Natural Science Foundation of China (NSFC) under grant No. 12075269. This work is also supported in part by the CAS-IHEP Fund for PRC$\backslash$US Collaboration in HEP. Radioassay work at PNNL was supported through a grant from the U.S. Department of Energy's Office of Nuclear Physics to support ton-scale neutrinoless double beta decay experimental Research and Development. PNNL is a multiprogram national laboratory operated by Battelle for the U.S. Department of Energy under Contract No. DE-AC05-76RL01830.

\bibliographystyle{IEEEtran}
\bibliography{IEEEabrv,Bibliography}

\begin{thebibliography}{10}
\providecommand{\url}[1]{#1}
\csname url@samestyle\endcsname
\providecommand{\newblock}{\relax}
\providecommand{\bibinfo}[2]{#2}
\providecommand{\BIBentrySTDinterwordspacing}{\spaceskip=0pt\relax}
\providecommand{\BIBentryALTinterwordstretchfactor}{4}
\providecommand{\BIBentryALTinterwordspacing}{\spaceskip=\fontdimen2\font plus
\BIBentryALTinterwordstretchfactor\fontdimen3\font minus
  \fontdimen4\font\relax}
\providecommand{\BIBforeignlanguage}[2]{{%
\expandafter\ifx\csname l@#1\endcsname\relax
\typeout{** WARNING: IEEEtran.bst: No hyphenation pattern has been}%
\typeout{** loaded for the language `#1'. Using the pattern for}%
\typeout{** the default language instead.}%
\else
\language=\csname l@#1\endcsname
\fi
#2}}
\providecommand{\BIBdecl}{\relax}
\BIBdecl

\bibitem{Heusser}
G.~Heusser, ``{LOW-RADIOACTIVITY BACKGROUND TECHNIQUES},'' \emph{Annu. Rev.
  Nucl. Part. Sci.}, vol.~45, pp. 543--590, 1995.

\bibitem{10.3389/fphy.2020.577734}
\BIBentryALTinterwordspacing
M.~Laubenstein and I.~Lawson, ``Low background radiation detection techniques
  and mitigation of radioactive backgrounds,'' \emph{Frontiers in Physics},
  vol.~8, 2020. [Online]. Available:
  \url{https://www.frontiersin.org/article/10.3389/fphy.2020.577734}
\BIBentrySTDinterwordspacing

\bibitem{Chepel:2012sj}
V.~Chepel and H.~Araujo, ``{Liquid noble gas detectors for low energy particle
  physics},'' \emph{JINST}, vol.~8, p. R04001, 2013.

\bibitem{MarrodanUndagoitia:2015veg}
T.~Marrod\'an~Undagoitia and L.~Rauch, ``{Dark matter direct-detection
  experiments},'' \emph{J. Phys. G}, vol.~43, no.~1, p. 013001, 2016.

\bibitem{Aprile:2009dv}
E.~Aprile and T.~Doke, ``{Liquid Xenon Detectors for Particle Physics and
  Astrophysics},'' \emph{Rev. Mod. Phys.}, vol.~82, pp. 2053--2097, 2010.

\bibitem{nEXO:2018ylp}
S.~A. Kharusi \emph{et~al.}, ``{nEXO Pre-Conceptual Design Report},'' 5 2018.

\bibitem{darkside}
{C.E. Aalsetch, et al.}, ``Darkside-20k: A 20 tonne two-phase lar tpc for
  direct dark matter detection at lngs,'' \emph{Eur. Phys. J. Plus}, vol. 133,
  p. 131, mar. 2018.

\bibitem{darwin}
{J. Aalbers, et al.}, ``Darwin: towards the ultimate dark matter detector,''
  \emph{Journal of Cosmology and Astroparticle Physics}, vol. 2016, no.~11, p.
  017, jun. 2016.

\bibitem{Avasthi:2021lgy}
A.~Avasthi \emph{et~al.}, ``{Kiloton-scale xenon detectors for neutrinoless
  double beta decay and other new physics searches},'' \emph{Phys. Rev. D},
  vol. 104, no.~11, p. 112007, 2021.

\bibitem{sipm}
D.~Renker, ``Geiger-mode avalanche photodiodes, history, properties and
  problems,'' \emph{Nucl. Instrum. Meth. A}, vol. 567, no.~1, pp. 48 --56, nov.
  2006.

\bibitem{nEXO:2021ujk}
G.~Adhikari \emph{et~al.}, ``{nEXO: neutrinoless double beta decay search
  beyond 10$^{28}$ year half-life sensitivity},'' \emph{J. Phys. G}, vol.~49,
  no.~1, p. 015104, 2022.

\bibitem{Kochanek:2020hmq}
I.~Kochanek, ``{Packaging strategies for large SiPM-based cryogenic
  photo-detectors},'' \emph{Nucl. Instrum. Meth. A}, vol. 980, p. 164487, 2020.

\bibitem{murayama2013warpage}
K.~Murayama, M.~Aizawa, K.~Hara, M.~Sunohara, K.~Miyairi, K.~Mori,
  J.~Charbonnier, M.~Assous, J.-P. Bally, G.~Simon \emph{et~al.}, ``Warpage
  control of silicon interposer for 2.5 d package application,'' in \emph{2013
  IEEE 63rd Electronic Components and Technology Conference}.\hskip 1em plus
  0.5em minus 0.4em\relax IEEE, 2013, pp. 879--884.

\bibitem{kapton}
I.~Arnquist, C.~Beck, M.~Vacri, K.~Harouaka, and R.~Saldanha, ``Ultra-low
  radioactivity kapton and copper-kapton laminates,'' \emph{Nuclear Instruments
  and Methods in Physics Research Section A: Accelerators, Spectrometers,
  Detectors and Associated Equipment}, vol. 959, p. 163573, 02 2020.

\bibitem{usman2017interposer}
A.~Usman, E.~Shah, N.~B. Satishprasad, J.~Chen, S.~A. Bohlemann, S.~H. Shami,
  A.~A. Eftekhar, and A.~Adibi, ``Interposer technologies for high-performance
  applications,'' \emph{IEEE Transactions on Components, Packaging and
  Manufacturing Technology}, vol.~7, no.~6, pp. 819--828, 2017.

\bibitem{banijamali2011advanced}
B.~Banijamali, S.~Ramalingam, K.~Nagarajan, and R.~Chaware, ``Advanced
  reliability study of tsv interposers and interconnects for the 28nm
  technology fpga,'' in \emph{2011 IEEE 61st Electronic Components and
  Technology Conference (ECTC)}.\hskip 1em plus 0.5em minus 0.4em\relax Ieee,
  2011, pp. 285--290.

\bibitem{lee2016overview}
C.-C. Lee, C.~Hung, C.~Cheung, P.-F. Yang, C.-L. Kao, D.-L. Chen, M.-K. Shih,
  C.-L.~C. Chien, Y.-H. Hsiao, L.-C. Chen \emph{et~al.}, ``An overview of the
  development of a gpu with integrated hbm on silicon interposer,'' in
  \emph{2016 IEEE 66th Electronic Components and Technology Conference
  (ECTC)}.\hskip 1em plus 0.5em minus 0.4em\relax IEEE, 2016, pp. 1439--1444.

\bibitem{khan2008development}
N.~Khan, S.~W. Yoon, A.~G. Viswanath, V.~Ganesh, R.~Nagarajan, D.~Witarsa,
  S.~Lim, and K.~Vaidyanathan, ``Development of 3-d stack package using silicon
  interposer for high-power application,'' \emph{IEEE transactions on advanced
  packaging}, vol.~31, no.~1, pp. 44--50, 2008.

\bibitem{ako_talk}
A.~Jamil, ``Large area sipms for ton-scale 0$\nu\beta\beta$,'' 2021, cPAD
  Instrumentation Frontier Workshop 2021.

\bibitem{velenis2009impact}
D.~Velenis, M.~Stucchi, E.~J. Marinissen, B.~Swinnen, and E.~Beyne, ``Impact of
  3d design choices on manufacturing cost,'' in \emph{2009 IEEE International
  Conference on 3D System Integration}.\hskip 1em plus 0.5em minus 0.4em\relax
  IEEE, 2009, pp. 1--5.

\bibitem{crook2013dielectric}
K.~Crook, M.~Carruthers, D.~Archard, S.~Burgess, and K.~Buchanan, ``Dielectric
  stack engineering for via-reveal passivation,'' in \emph{2013 IEEE 63rd
  Electronic Components and Technology Conference}.\hskip 1em plus 0.5em minus
  0.4em\relax IEEE, 2013, pp. 576--580.

\bibitem{huang2013integration}
B.~K. Huang, C.~M. Lin, S.~J. Huang, C.~W. Chiang, P.~C. Huang, G.~X. Chen,
  C.~C. Chao, and C.~H. Lu, ``Integration challenges of tsv backside via reveal
  process,'' in \emph{2013 IEEE 63rd Electronic Components and Technology
  Conference}.\hskip 1em plus 0.5em minus 0.4em\relax IEEE, 2013, pp. 915--917.

\bibitem{kumar2012robust}
N.~Kumar, S.~Ramaswami, J.~Dukovic, J.~Tseng, R.~Ding, N.~Rajagopalan,
  B.~Eaton, R.~Mishra, R.~Yalamanchili, Z.~Wang \emph{et~al.}, ``Robust tsv
  via-middle and via-reveal process integration accomplished through
  characterization and management of sources of variation,'' in \emph{2012 IEEE
  62nd Electronic Components and Technology Conference}.\hskip 1em plus 0.5em
  minus 0.4em\relax IEEE, 2012, pp. 787--793.

\bibitem{D0JA00220H}
\BIBentryALTinterwordspacing
K.~Harouaka, E.~W. Hoppe, and I.~J. Arnquist, ``A novel method for measuring
  ultra-trace levels of u and th in au{,} pt{,} ir{,} and w matrices using
  icp-qqq-ms employing an o2 reaction gas,'' \emph{J. Anal. At. Spectrom.},
  vol.~35, pp. 2859--2866, 2020. [Online]. Available:
  \url{http://dx.doi.org/10.1039/D0JA00220H}
\BIBentrySTDinterwordspacing

\bibitem{ARNQUIST2020163573}
\BIBentryALTinterwordspacing
I.~J. Arnquist, C.~Beck, M.~L. {di Vacri}, K.~Harouaka, and R.~Saldanha,
  ``Ultra-low radioactivity kapton and copper-kapton laminates,'' \emph{Nuclear
  Instruments and Methods in Physics Research Section A: Accelerators,
  Spectrometers, Detectors and Associated Equipment}, vol. 959, p. 163573,
  2020. [Online]. Available:
  \url{https://www.sciencedirect.com/science/article/pii/S0168900220301480}
\BIBentrySTDinterwordspacing

\end{thebibliography}

\end{document}